\documentclass[10pt,a4paper]{article}
\usepackage[utf8]{inputenc}
\usepackage[english]{babel}
\usepackage{amsmath}
\usepackage{amsfonts}
\usepackage{amssymb}
\usepackage{amsthm}
\usepackage{mathrsfs}
\usepackage{dsfont}
\usepackage[left=2cm,right=2cm,top=2cm,bottom=2cm]{geometry}
\usepackage[dvipsnames]{xcolor}

\usepackage{tikz}
\usetikzlibrary{arrows, arrows.meta, angles, positioning, quotes, shapes, matrix, calc, plotmarks, pgfplots.groupplots}
\usepackage{pgfplots}
\pgfplotsset{compat=newest}
\usepackage{pgfplotstable}
\usepgfplotslibrary{fillbetween}

\pgfplotsset{every tick label/.append style={font=\footnotesize}}

\usepackage{algorithm}
\usepackage{algpseudocode}

\PassOptionsToPackage{hyphens}{url}
\RequirePackage[colorlinks,citecolor=blue,urlcolor=blue]{hyperref}

\newcommand{\base}{./data_for_tex}

\theoremstyle{remark}

\numberwithin{equation}{section}

\providecommand{\keywords}[1]
{
	\textbf{\textit{Keywords--}} #1
}

\DeclareMathOperator{\relu}{ReLU}
\DeclareMathOperator{\e}{e}
\DeclareMathOperator{\MS}{MS}
\DeclareMathOperator{\diag}{diag}

\title{Langevin algorithms for Markovian Neural Networks and Deep Stochastic control}
\author{Pierre Bras\footnote{Sorbonne Universit\'e, Laboratoire de Probabilit\'es, Statistique et Mod\'elisation, UMR 8001, case 158, 4 pl. Jussieu, F-75252 Paris Cedex 5, France. E-mail: \texttt{pierre.bras@sorbonne-universite.fr} and \texttt{gilles.pages@sorbonne-universite.fr}.} \footnote{Corresponding author.} $\ $ and Gilles Pag\`es\footnotemark[1]}
\date{}

\begin{document}

\maketitle

\begin{abstract}
Stochastic Gradient Descent Langevin Dynamics (SGLD) algorithms, which add noise to the classic gradient descent, are known to improve the training of neural networks in some cases where the neural network is very deep.
In this paper we study the possibilities of training acceleration for the numerical resolution of stochastic control problems through gradient descent, where the control is parametrized by a neural network. If the control is applied at many discretization times then solving the stochastic control problem reduces to minimizing the loss of a very deep neural network.
We numerically show that Langevin algorithms improve the training on various stochastic control problems like hedging and resource management, and for different choices of gradient descent methods.
\end{abstract}


\section{Introduction}

Stochastic Optimal Control (SOC), which consists in optimizing a functional of a trajectory of a controlled Stochastic Differential Equation (SDE) has applications in a wide range of problems: management of resources, queuing systems, epidemic and population processes, pricing of financial derivatives, portfolio allocation...
In comparison with classic optimal control, SOC models include a random noise with known probability distribution that affects the evolution or the observation of the system. SOC also aims at managing the risk induced by this noise.

SOC problems are usually solved using specific strategies, such as Forward-Backward SDEs (FBSDEs) \cite{peng1999}, or by solving Hamilton-Jacobi-Bellman (HJB) optimality conditions \cite{bellman2010} through partial differential equations methods using appropriate numerical schemes or by stochastic dynamic programming \cite{kushner2001}. Such problems can also be solved using Neural Networks calibrated by SGD techniques \cite{gobet2005,han2016,wang2019,carmona2021,bachouch2022}.

More specifically, in this article we consider the numerical resolution of a SOC problem where the control is parametrized by a neural network calibrated by gradient descent.
This method implies to compute the path-wise derivatives along the trajectory of the SDE of the objective function with respect to the parameters of the neural network, as introduced in \cite{giles2005,giles2007}.
Stochastic gradient descent is a very general approach that can be applied to a wide range of problems and which does not need to be specifically adapted to each problem under consideration. Moreover, SGD scales very efficiently to high-dimensional problems, in contrast with HJB-based methods, and has proved its efficiency on highly non-convex problems \cite{dauphin2015}.

\medskip

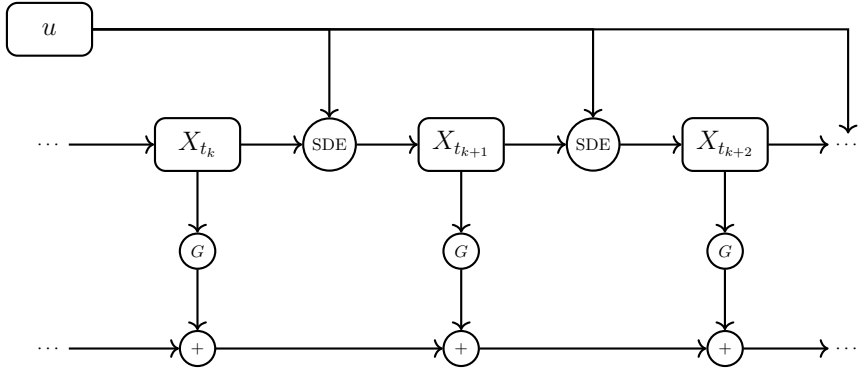
\begin{figure}
\centering

\begin{tikzpicture}[
   thick,scale=0.7, every node/.style={scale=0.7},
   box/.style = {draw, rounded corners, 
                 minimum width=16mm, minimum height=10mm, align=center, font=\Large},
   circ/.style =  {draw, circle, minimum width=2mm, minimum height=2mm, align=center, font=\small},
         auto = right,]
         
\matrix[row sep=0.8cm,column sep=0.8cm] {
    \node [box](u) {$u$};   & & & & & & \\
    \node (in) {$\cdots$}; & \node [box](X0) {$X_{t_k}$}; & \node [circ](SDE1) {SDE}; & \node [box](X1) {$X_{t_{k+1}}$}; & \node [circ](SDE2) {SDE}; & \node [box](X2) {$X_{t_{k+2}}$}; & \node (out) {$\cdots$};   \\
    & \node [circ](G0) {$G$}; & & \node [circ](G1) {$G$}; & & \node [circ](G2) {$G$}; & \\
    \node (plus_in) {$\cdots$}; & \node [circ](plus0) {$+$}; & & \node [circ](plus1) {$+$}; & & \node [circ](plus2) {$+$}; & \node (plus_out) {$\cdots$}; \\
};         
        
\draw[->] (in) -- (X0);
\draw[->] (X0) -- (SDE1);
\draw[->] (SDE1) -- (X1);
\draw[->] (X1) -- (SDE2);
\draw[->] (SDE2) -- (X2);
\draw[->] (X2) -- (out);

\draw[->] (u) -| (SDE1);
\draw[->] (u) -| (SDE2);
\draw[->] (u) -| (out);

\draw[->] (X0) -- (G0);
\draw[->] (G0) -- (plus0);
\draw[->] (X1) -- (G1);
\draw[->] (G1) -- (plus1);
\draw[->] (X2) -- (G2);
\draw[->] (G2) -- (plus2);

\draw[->] (plus_in) -- (plus0);
\draw[->] (plus0) -- (plus1);
\draw[->] (plus1) -- (plus2);
\draw[->] (plus2) -- (plus_out);
         
\end{tikzpicture}

\caption{Depth of Markovian neural networks. The control $u$ acts on $X_{t_k}$, which itself acts on $X_{t_{k+1}}$, $X_{t_{k+2}}$, $\ldots$, $X_{t_N}$, hence the depth of the network.}
\label{fig:schema:1}
\end{figure}

\begin{figure}
\centering

\begin{tikzpicture}[
   thick,scale=0.7, every node/.style={scale=0.7},
   box/.style = {draw, rounded corners, 
                 minimum width=16mm, minimum height=10mm, align=center, font=\Large},
   circ/.style =  {draw, circle, minimum width=2mm, minimum height=2mm, align=center, font=\small},
         auto = right,]
         
\matrix[row sep=0.8cm,column sep=0.8cm] {
     & & \node [box](u0) {$u_{t_k}$}; & & \node [box](u1) {$u_{t_{k+1}}$}; & & \\
    \node (in) {$\cdots$}; & \node [box](X0) {$X_{t_k}$}; & \node [circ](SDE1) {SDE}; & \node [box](X1) {$X_{t_{k+1}}$}; & \node [circ](SDE2) {SDE}; & \node [box](X2) {$X_{t_{k+2}}$}; & \node (out) {$\cdots$};   \\
    & \node [circ](G0) {$G$}; & & \node [circ](G1) {$G$}; & & \node [circ](G2) {$G$}; & \\
    \node (plus_in) {$\cdots$}; & \node [circ](plus0) {$+$}; & & \node [circ](plus1) {$+$}; & & \node [circ](plus2) {$+$}; & \node (plus_out) {$\cdots$}; \\
};         
        
\draw[->] (in) -- (X0);
\draw[->] (X0) -- (SDE1);
\draw[->] (SDE1) -- (X1);
\draw[->] (X1) -- (SDE2);
\draw[->] (SDE2) -- (X2);
\draw[->] (X2) -- (out);

\draw[->] (u0) -- (SDE1);
\draw[->] (u1) -- (SDE2);

\draw[->] (X0) -- (G0);
\draw[->] (G0) -- (plus0);
\draw[->] (X1) -- (G1);
\draw[->] (G1) -- (plus1);
\draw[->] (X2) -- (G2);
\draw[->] (G2) -- (plus2);

\draw[->] (plus_in) -- (plus0);
\draw[->] (plus0) -- (plus1);
\draw[->] (plus1) -- (plus2);
\draw[->] (plus2) -- (plus_out);
         
\end{tikzpicture}

\caption{Markovian neural network with one control for every time step.}
\label{fig:schema:2}
\end{figure}
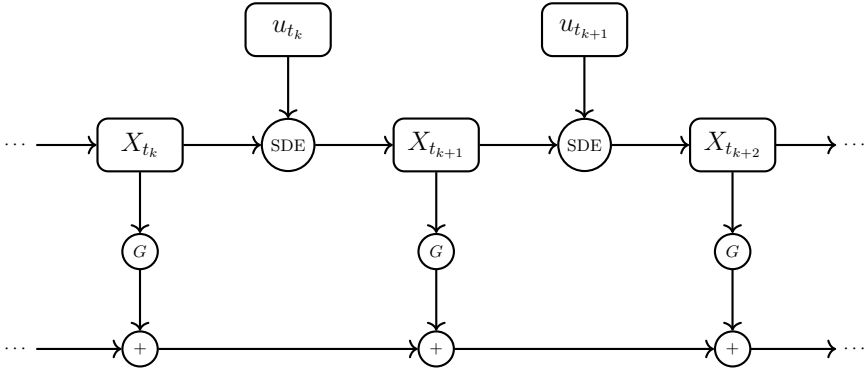

However, if the neural control is applied at many time steps as it is the case for the Euler-Maruyama scheme where the (discrete) control is taken as an approximation of a continuous control, then the SOC problem reads as the optimization of a very deep neural network, which is roughly as deep as the number of instants at which the control can be applied (see Figures \ref{fig:schema:1} and \ref{fig:schema:2}).
Very deep neural networks are known to be considerably more difficult to train \cite{glorot2010,dauphin2014} and may run into vanishing gradient problems \cite{hochreiter1991,hanin2018}. Indeed, the deeper the neural network is, the more non-linear it is, thus increasing the number of local traps for the gradient descent such as local (but not global) and saddle points.
In image analysis where very deep convolutional neural networks are commonly used, residual \cite{residual} and convolutional dense \cite{huang2017} networks were introduced to deal with this issue. These networks are based on architectures with residual connections to propagate the gradient information through the numerous successive layers.

As it comes to deep SOC, we cannot freely change the structure of the neural network since it is fixed by the equations defining the SOC problem and therefore we cannot directly use residual connections. We can only freely choose the structure of the neural network returning the control, for which a few layers is often enough (see for example \cite{buehler2019,bachouch2022}).

A way to improve the learning is to replace SGD algorithms by Stochastic Gradient Langevin Dynamics (SGLD) algorithms. Such optimizers add an exogenous white noise to the gradient descent, providing regularization and allowing to escape from traps. It has indeed been observed that adding noise improves the learning for very deep neural networks \cite{neelakantan2015,asgld,gulcehre2016,shridhar2019}. Moreover, \cite{bras2022} compares side-by-side Langevin with non-Langevin algorithms on networks with increasing depth and shows that for shallow neural networks, Langevin algorithms do nothing else than adding noise to the gradient descent, however the deeper the network is, the greater the gains provided by Langevin algorithms are.

\medskip

In the present article we study the performances of Langevin optimizers on SOC problems where the number of discretization times where the control is applied is large enough.
We use the preconditioned versions of SGD and SGLD \cite{li2015} for various choices of preconditioners.
We compare side-by-side Langevin and non-Langevin algorithms and we show that Langevin optimizers can significantly improve the training procedure on various problems: fishing quotas \cite{lauriere2021}, deep hedging \cite{buehler2019}, oil drilling and resource management \cite{gaigi2021}.
We mainly consider two different approaches for numerical resolution of SOCs.
In the first approach, the control is a single neural network which is applied to every time step and which may depend on the running time $t$ (see Figure \ref{fig:schema:1}). This approach leads to a model with fewer trainable parameters, which is critical in some data-driven financial applications where the amount of data is limited, and which is more able to capture the specific Markovian features of the problem.
In the second approach, a different neural network is used for each control time (see Figure \ref{fig:schema:2}). This last setting is also suitable for applying the Layer Langevin algorithm, which is a variant of the Langevin algorithm introduced in \cite{bras2022} and which proved to be more adapted to the training of very deep neural networks than the Langevin algorithm itself.

We observe that the gains of Langevin algorithms depend on the preconditioner however. While the Adam \cite{adam} and the Adadelta \cite{adadelta} algorithms can be substantially accelerated by Langevin training, the gains are more limited or sometimes null for RMSprop \cite{rmsprop}.

\medskip

The code for the numerical experiments is available at
\url{https://github.com/Bras-P/langevin-for-stochastic-control}.
It includes in particular ready-to-use Langevin optimizers and Layer Langevin optimizers as instances of the TensorFlow \texttt{Optimizer} base class, a framework for algorithm comparison in a SOC setting with GPU support and a demonstration notebook.

\medskip

\textbf{Notations:}
For $x$ and $y$ two vectors we define the Schur product $x \ast y$ as the vector $(x_i y_i)_i$. We also sometimes use the notation $\ast$ for scalar-vector multiplication.
For $a,b \in \mathbb{N}$ we denote $\mathcal{M}_{a,b}(\mathbb{R})$ the set of $a \times b$ matrices with real-valued coefficients.
For $x \in \mathbb{R}$ we define the positive part of $x$ denoted $x_+$ as $\max(x,0)$.
We consider multivariate $(\mathcal{F}_t)$-Brownian motions $W$ and $B$ defined on some filtered probability space $(\Omega, \mathcal{F}, (\mathcal{F}_t)_{t \in [0,T]}), \mathbb{P})$.

\section{Stochastic control through gradient descent}

\subsection{Stochastic optimal control}

We consider the following SOC problem in continuous time:
\begin{align}
\label{eq:def_J}
& \min_{u} J(u) := \mathbb{E}\left[ \int_0^T G(t,X_t)dt + F(X_T) \right], \\
\label{eq:def_X}
& dX_t = b(X_t,u_t)dt + \sigma(X_t,u_t)dW_t, \ t \in [0,T]
\end{align}
where $b :\mathbb{R}^{d_1} \times \mathbb{R}^{d_3} \to \mathbb{R}^{d_1}$, $\sigma : \mathbb{R}^{d_1} \times \mathbb{R}^{d_3} \to \mathcal{M}_{d_1,d_2}(\mathbb{R})$, $W$ is a $\mathbb{R}^{d_2}$-valued Brownian motion and $u$ is a $\mathbb{R}^{d_3}$-valued continuous adapted process, $T>0$, $G:[0,T] \times \mathbb{R}^{d_1} \to \mathbb{R}$ and $F:\mathbb{R}^{d_1} \to \mathbb{R}$.

We first approximate the continuous SDE $(X_t)_{t \in [0,T]}$ with its Euler-Maruyama scheme and the control $u_t$ with a discrete-time control. For $N \in \mathbb{N}$ being the number of discretization times, we consider the regular subdivision of $[0,T]$:
\begin{equation}
t_k := kT/N, \ k \in \lbrace 0, \ldots, N \rbrace, \quad h:= T/N
\end{equation}
and we approximate the control applied at times $t_0, \ldots, t_{N-1}$ as the output of a single neural network depending on $t$, or as the output of $N$ neural networks, one for each discretization instant $t_k$:
\begin{equation}
\label{eq:def_u}
u_{t_k} = \bar{u}_\theta(t_k, X_{t_k}) \quad \text{or} \quad u_{t_k} = \bar{u}_{\theta^k}(X_{t_k}) 
\end{equation}
where $\bar{u}_{\theta}$ is a neural function with finite-dimensional parameter $\theta \in \mathbb{R}^{d}$. Indeed, since \eqref{eq:def_X} defines a Markovian process, we can assume that $u_t$ depends only on $t$ and on $X_t$ instead of $t$ and $(X_s)_{s \in [0,t]}$.

The SOC problem \eqref{eq:def_J} is numerically approximated by:
\begin{align}
\label{eq:def_J_bar}
& \min_{\theta} \bar{J}(\bar{u}_\theta) := \sum_{k=0}^{N-1} (t_{k+1}-t_k) G(t_{k+1}, \bar{X}^\theta_{t_{k+1}}) + F(\bar{X}^\theta_{t_N}), \\
\label{eq:X_t_k}
& \bar{X}^\theta_{t_{k+1}} = \bar{X}^\theta_{t_k} + (t_{k+1}-t_k) b\big(\bar{X}^\theta_{t_k}, \bar{u}_{k,\theta}(\bar{X}^\theta_{t_k})\big) + \sqrt{t_{k+1}-t_k} \sigma\big(\bar{X}^\theta_{t_k}, \bar{u}_{k,\theta}(\bar{X}^\theta_{t_k})\big) \xi_{k+1}, \\
& \xi_k \underset{\textup{i.i.d.}}{\sim} \mathcal{N}(0,I_{d_2})
\end{align}
where $\theta \in \mathbb{R}^d$ and $\bar{u}_{k,\theta} = \bar{u}_\theta(t_k,\cdot)$ in the first case of \eqref{eq:def_u} and $\theta = (\theta^0, \ldots, \theta^{N-1}) \in (\mathbb{R}^d)^N$ and $\bar{u}_{k,\theta} = \bar{u}_{\theta^k}$ in the second case of \eqref{eq:def_u}.

For every $\theta$, $\nabla_\theta \bar{J}$ can be computed by automatic differentiation as the gradient w.r.t. to $\theta$ is tracked all along the trajectory through the recursive relation \eqref{eq:X_t_k} \cite{giles2005,giles2007}. Then the SGD algorithm reads
\begin{equation}
\label{eq:SGD_J}
\theta_{n+1} = \theta_n - \gamma_{n+1} \frac{1}{n_{\text{batch}}} \sum_{i=1}^{n_{\text{batch}}} \nabla_\theta \bar{J}(\bar{u}_{\theta_n}, (\xi^{i,n+1}_k)_{1 \le k \le N}) =: \theta_n - \gamma_{n+1} g_{n+1}
\end{equation}
where $(\xi^{i,n}_k)_{1 \le k \le N, 1 \le i \le n_{\text{batch}}, n \in \mathbb{N}}$ is an array of i.i.d. random vectors $\mathcal{N}(0,I_{d_2})$-distributed, $(\gamma_n)_{n \in \mathbb{N}}$ is a non-increasing positive step sequence and where the dependence of $\bar{J}$ in $(\xi^{i,n}_k)$ is made explicit.

If the number of Euler-Maruyama steps $N$ is large, then the optimization problem in \eqref{eq:def_J_bar} consists in the training of a very deep neural network that can be difficult to train directly (see the Introduction).
Both cases are illustrated in Figures \ref{fig:schema:1} and \ref{fig:schema:2}.

\subsection{Preconditioned stochastic gradient Langevin dynamics}

We consider preconditioned stochastic gradient algorithms i.e. for $(P_n)$ a preconditioner rule the update reads
\begin{equation}
\label{eq:psgd_def}
\theta_{n+1} = \theta_n - \gamma_{n+1} P_{n+1} \cdot g_{n+1}
\end{equation}
where $g_{n+1}$ is defined in \eqref{eq:SGD_J}.
We use the Adam \cite{adam}, the RMSprop \cite{rmsprop} and the Adadelta \cite{adadelta} preconditioners, which are detailed in Algorithms \ref{algo:adam}, \ref{algo:rms_prop} and \ref{algo:adadelta} respectively.
For some algorithm \textit{name}, the corresponding Langevin algorithm denoted L-\textit{name} reads
\begin{equation}
\label{eq:sgld_def}
\theta_{n+1} = \theta_n - \gamma_{n+1} P_{n+1} \cdot g_{n+1} + \sigma_{n+1} \sqrt{\gamma_{n+1}} \mathcal{N}(0,P_{n+1})
\end{equation}
where $(\sigma_n)$ is a constant or non-decreasing sequence controlling the amount of injected noise.

\begin{minipage}{0.46\textwidth}
\begin{algorithm}[H]
\caption{Adam update}\label{algo:adam}
\begin{algorithmic}
\State \textbf{Parameters:} $\beta_1, \beta_2, \lambda > 0$
\State $M_{n+1} = \beta_1 M_n + (1-\beta_1) g_{n+1} $
\State $\MS_{n+1} = \beta_2 \MS_n + (1-\beta_2) g_{n+1} \odot g_{n+1} $
\State $\widehat{M}_{n+1} = M_{n+1}/(1-\beta_1^{n+1}) $
\State $\widehat{\MS}_{n+1} = \MS_{n+1}/(1-\beta_2^{n+1}) $
\State $P_{n+1} = \diag\big( \mathds{1} \oslash \big(\lambda \mathds{1} + \sqrt{\widehat{\MS}_{n+1}}\big) \big) $
\State $\theta_{n+1} = \theta_n - \gamma_{n+1} P_{n+1} \cdot \widehat{M}_{n+1} .$
\end{algorithmic}
\end{algorithm}
\end{minipage}
\hfill
\begin{minipage}{0.46\textwidth}
\begin{algorithm}[H]
\caption{RMSprop update}\label{algo:rms_prop}
\begin{algorithmic}
\State \textbf{Parameters:} $\alpha, \lambda > 0$
\State $\MS_{n+1} = \alpha \MS_n + (1-\alpha) g_{n+1} \odot g_{n+1}$
\State $P_{n+1} = \diag\left( \mathds{1} \oslash \left(\lambda \mathds{1} + \sqrt{\MS_{n+1}}\right) \right)$
\State $\theta_{n+1} = \theta_n - \gamma_{n+1} P_{n+1} \cdot g_{n+1}$
\end{algorithmic}
\end{algorithm}
\end{minipage}
~
\begin{algorithm}
\caption{Adadelta update}\label{algo:adadelta}
\begin{algorithmic}
\State \textbf{Parameters:} $\beta_1, \beta_2, \lambda > 0$
\State $\MS_{n+1} = \beta_1 \MS_n + (1-\beta_1) g_{n+1} \odot g_{n+1} $
\State $P_{n+1} = \diag\big( (\lambda \mathds{1} + \widehat{\MS}_n) \oslash \big(\lambda \mathds{1} + \sqrt{\widehat{\MS}_{n}}\big) \big) $
\State $\theta_{n+1} = \theta_n - \gamma_{n+1} P_{n+1} \cdot g_{n+1}.$
\State $\widehat{\MS}_{n+1} = \beta_2 \MS_n + (1-\beta_2) (\theta_{n+1}-\theta_n) \odot (\theta_{n+1}-\theta_n)$.
\end{algorithmic}
\end{algorithm}

The Layer Langevin algorithm, introduced in \cite{bras2022} consists in updating with Langevin noise only some layers of the network. It relies on the heuristic that for a deep neural network, the non-linearities of the network are mostly contained in the deepest layers and adds Langevin noise to these layers only.

Choosing a preconditioner rule $P$ called \textit{name}, the Layer Langevin algorithm denoted LL-\textit{name} reads
\begin{align}
\label{eq:layer_langevin}
\theta^{(i)}_{n+1} & = \theta^{(i)}_n - \gamma_{n+1} [P_{n+1} \cdot g_{n+1}]^{(i)} + \mathds{1}_{i \in \mathcal{J}} \sigma_{n+1} \sqrt{\gamma_{n+1}} \big[\mathcal{N}(0, P_{n+1}) \big]^{(i)},
\end{align}
where $\mathcal{J}$ is a subset of weight indices.
In particular, we denote LL-\textit{name} $p\%$ the Layer Langevin \textit{name} algorithm where the Langevin layers are chosen to be the first $p\%$ layers.

\subsection{Experimental setting}

We proceed to side-by-side comparison of Langevin algorithms \eqref{eq:sgld_def} with their non-Langevin counterparts \eqref{eq:psgd_def} on various SOC problems.

We consider a first case where the control is given by only one neural network depending on $t$ and a second case where a different neural network is used for each control time \eqref{eq:def_u}.
In this second case, since we can expect two consecutive control networks to have close parameters, one usual way of performing the training procedure is to first train networks for a small amount of control times, then to perform the whole training through transfer learning.
We do not expect Langevin algorithms to be suitable for the fine tuning, but since the first step of the training still consists in training a deep neural network, we analyse the benefits of Langevin algorithms for this first step.

Unless stated otherwise, the batch size is set to $n_{\text{batch}}=512$ i.e. stochastic gradient iterations are performed by averaging the gradient over 512 random trajectories. In the plot, each epoch consists in 5 batches i.e. the average loss $J(u_\theta)$ is plotted every 5 iterations of the stochastic gradient. After each epoch, $J(u_\theta)$ is estimated over $25 \times 512$ trajectories and 95\% confidence intervals are indicated, although for some plots the intervals are too small to be visible.
While comparing some algorithm with its Langevin or Layer Langevin counterpart, we ensure that both training procedures start with the same initial weights.

\section{Fishing quotas}

We consider the fishing quota problem introduced in \cite{lauriere2021}. Let $X_t \in \mathbb{R}^{d_1}$ be the fish biomass for every fish species; we wish to keep it close to an ideal state $\mathcal{X}_t \in \mathbb{R}^{d_1}$. The dynamics of $X$ are given by
\begin{equation}
dX_t = X_t \ast \left( (r - u_t - \kappa X_t)dt + \eta dW_t \right), \ t \in [0,T],
\end{equation}
where $r \in \mathbb{R}^{d_1}$ is the growth rate for each species, $u_t \in \mathbb{R}^{d_1}$ is the controlled fishing (with $d_3=d_1$), $\kappa \in \mathcal{M}_{d_1,d_1}(\mathbb{R})$ is the interaction matrix between the fish species, $\eta \in \mathcal{M}_{d_1,d_2}(\mathbb{R})$, $W$ is a $\mathbb{R}^{d_2}$-valued Brownian motion. The control $u$ is constrained to take its values in $[u_m, u_M]^{d_1}$. The objective is
\begin{equation}
J(u) = \mathbb{E} \left[ \int_0^T (|X_t - \mathcal{X}_t|^2 - \langle \alpha, u_t \rangle) dt + \beta [u]^{0,T} \right],
\end{equation}
where $\alpha \in \mathbb{R}^{d_1}$, $\beta \in \mathbb{R}^+$, $[u]^{0,T}$ denotes the quadratic variation of $u$ on $[0,T]$. The term $\langle \alpha, u \rangle$ penalizes small fishing quotas while the term $\beta [u]^{0,T}$ penalizes too many daily changes.

\smallskip

In the experiments, following \cite{lauriere2021} we choose
\begin{align}
d_1 = d_2 = 5, \ T=1,\ \mathcal{X} \equiv \mathbf{1}, \ r = 2 \ast \mathbf{1}, \ \eta = 0.1 \ast I_{d_1}, \ \alpha=0.01 \ast\mathbf{1}, \ \beta=0.1, \ u_m=0.1, \ u_M=1 .
\end{align}
and \begin{equation}
\kappa = \begin{pmatrix}
1.2 & -0.1 & 0 & 0 & -0.1 \\
0.2 & 1.2 & 0 & 0 & -0.1 \\
0 & 0.2 & 1.2 & -0.1 & 0 \\
0 & 0 & 0.1 & 1.2 & 0 \\
0.1 & 0.1 & 0 & 0 & 1.2
\end{pmatrix}.
\end{equation}
The initial state $X_0$ is randomly generated following $\mathcal{N}\big(\mathbf{1}, (1/2)I_{d_1}\big)$ clipped to $[0.2,2]^{d_1}$.
The quadratic variation $[u]^{0,T}$ is approximated in the discretized setting by
\begin{equation}
[u]^{0,T} \simeq \sum_{k=0}^{N-1} |u_{t_{k+1}} - u_{t_k}|^2.
\end{equation}
Each control $u_\theta$ is given by a feedforward neural network with two hidden layers with $32$ units each and with $\relu$ activation while the output layer has sigmoid activation in order to fulfil the constraint on $u$.
An example of controlled trajectory is plotted in Figure \ref{fig:fishing:traj}.

\begin{figure}
\centering

\begin{tikzpicture}
	\begin{groupplot}[group style={
					group size=2 by 1,
					horizontal sep=5ex,
					x descriptions at=edge bottom,
					group name=plots,
					},
			  xlabel=Epochs,
			  ytick pos=left,
			  height=0.25\textheight,
			  width=0.5\textwidth,
			  every axis plot/.append style={line width=1pt,mark size=0.8pt,mark options={solid}},
			  xmin=0,xmax=50,
			  grid=both,
			  minor grid style={gray!25},
			  major grid style={gray!25},
			  ]

		\nextgroupplot[ylabel=$J(u_\theta)$, ymax=0.6]
\addplot[color=black, mark=*]
	table[x=time,y=f,col sep=comma]{\base/fishing/rmsprop_N50/0_v_loss.csv};
\addlegendentry{RMSprop};
	
\addplot[color=blue, mark=square*, style=densely dotted]
	table[x=time,y=f,col sep=comma]{\base/fishing/rmsprop_N50/1_v_loss.csv};
\addlegendentry{L-RMSprop};
	
	\addplot[draw=none, name path=0_plus] table[x=time,y=f_plus,col sep=comma]{\base/fishing/rmsprop_N50/0_v_loss.csv};
	\addplot[draw=none, name path=0_minus] table[x=time,y=f_minus,col sep=comma]{\base/fishing/rmsprop_N50/0_v_loss.csv};
	\addplot[black!30] fill between[of=0_minus and 0_plus];

	\addplot[draw=none, name path=1_plus] table[x=time,y=f_plus,col sep=comma]{\base/fishing/rmsprop_N50/1_v_loss.csv};
	\addplot[draw=none, name path=1_minus] table[x=time,y=f_minus,col sep=comma]{\base/fishing/rmsprop_N50/1_v_loss.csv};
	\addplot[blue!30] fill between[of=1_minus and 1_plus];

		\nextgroupplot[]
\addplot[color=black, mark=*]
	table[x=time,y=f,col sep=comma]{\base/fishing/adadelta_N50/0_v_loss.csv};
\addlegendentry{Adadelta};

\addplot[color=blue, mark=square*, style=densely dotted]
	table[x=time,y=f,col sep=comma]{\base/fishing/adadelta_N50/1_v_loss.csv};
\addlegendentry{L-Adadelta};
	
	\addplot[draw=none, name path=0_plus] table[x=time,y=f_plus,col sep=comma]{\base/fishing/adadelta_N50/0_v_loss.csv};
	\addplot[draw=none, name path=0_minus] table[x=time,y=f_minus,col sep=comma]{\base/fishing/adadelta_N50/0_v_loss.csv};
	\addplot[black!30] fill between[of=0_minus and 0_plus];
	
	\addplot[draw=none, name path=1_plus] table[x=time,y=f_plus,col sep=comma]{\base/fishing/adadelta_N50/1_v_loss.csv};
	\addplot[draw=none, name path=1_minus] table[x=time,y=f_minus,col sep=comma]{\base/fishing/adadelta_N50/1_v_loss.csv};
	\addplot[blue!30] fill between[of=1_minus and 1_plus];
	
	\end{groupplot}
\end{tikzpicture}

\begin{tikzpicture}
\begin{axis}[
	ylabel=Zoom on RMSp.,
	label style={font=\tiny},
	xmin=20,
	ymax=0.42,
	grid=both,
	minor grid style={gray!25},
	major grid style={gray!25},
	width=0.5\linewidth,
	height=0.12\paperheight,
	line width=1pt,
	mark size=0.8pt,
	mark options={solid},
	]
\addplot[color=black, mark=*]
	table[x=time,y=f,col sep=comma]{\base/fishing/rmsprop_N50/0_v_loss.csv};
	
\addplot[color=blue, mark=square*, style=densely dotted]
	table[x=time,y=f,col sep=comma]{\base/fishing/rmsprop_N50/1_v_loss.csv};
	
	\addplot[draw=none, name path=0_plus] table[x=time,y=f_plus,col sep=comma]{\base/fishing/rmsprop_N50/0_v_loss.csv};
	\addplot[draw=none, name path=0_minus] table[x=time,y=f_minus,col sep=comma]{\base/fishing/rmsprop_N50/0_v_loss.csv};
	\addplot[black!30] fill between[of=0_minus and 0_plus];

	\addplot[draw=none, name path=1_plus] table[x=time,y=f_plus,col sep=comma]{\base/fishing/rmsprop_N50/1_v_loss.csv};
	\addplot[draw=none, name path=1_minus] table[x=time,y=f_minus,col sep=comma]{\base/fishing/rmsprop_N50/1_v_loss.csv};
	\addplot[blue!50,opacity=0.5,] fill between[of=1_minus and 1_plus];
\end{axis}
\end{tikzpicture}

\caption{Comparison of Langevin algorithms with their non-Langevin counterparts during the training for the fishing control problem with $N=50$ respectively.  The schedules are $\gamma_n=2\e-3$ ($5\e-1$) and $\sigma_n=5\e-3$ ($1\e-2$) for RMSprop (Adadelta resp.) for epochs 0 to 40 and $\gamma_n$ is divided by 10 and $\sigma_n$ is set to 0 beyond. At the end of each epoch, $J$ is estimated over $50 \times 512$ trajectories. A zoom on the last epochs for RMSprop is given.}
\label{fig:fishing:2}
\end{figure}
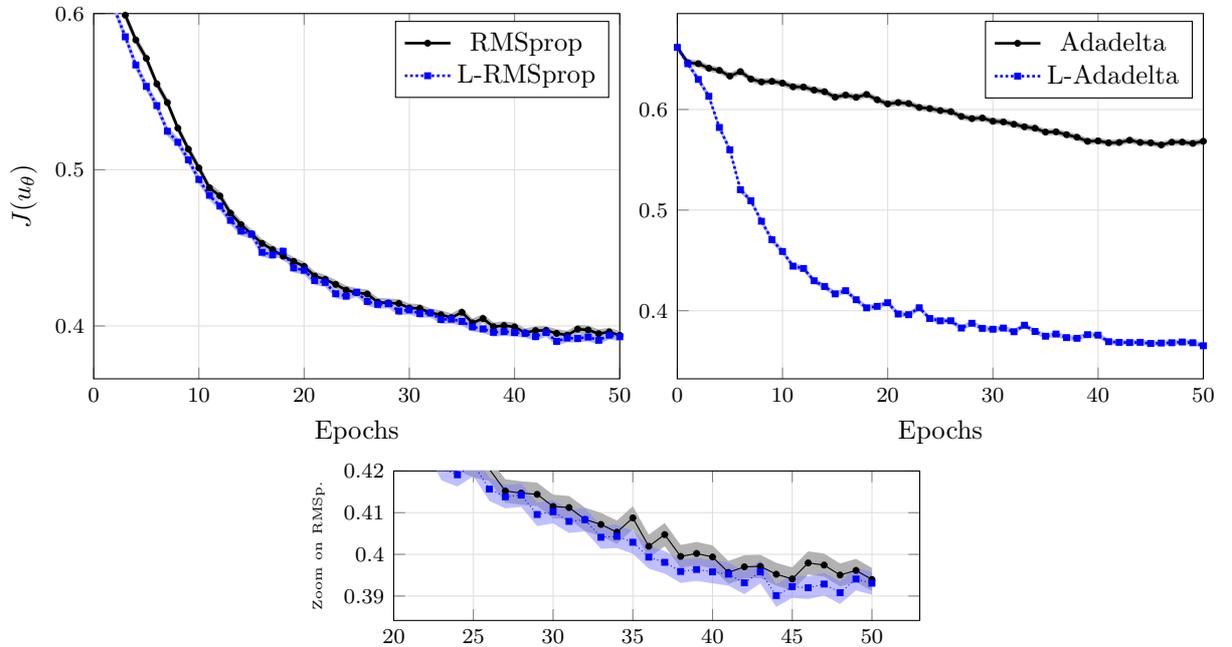

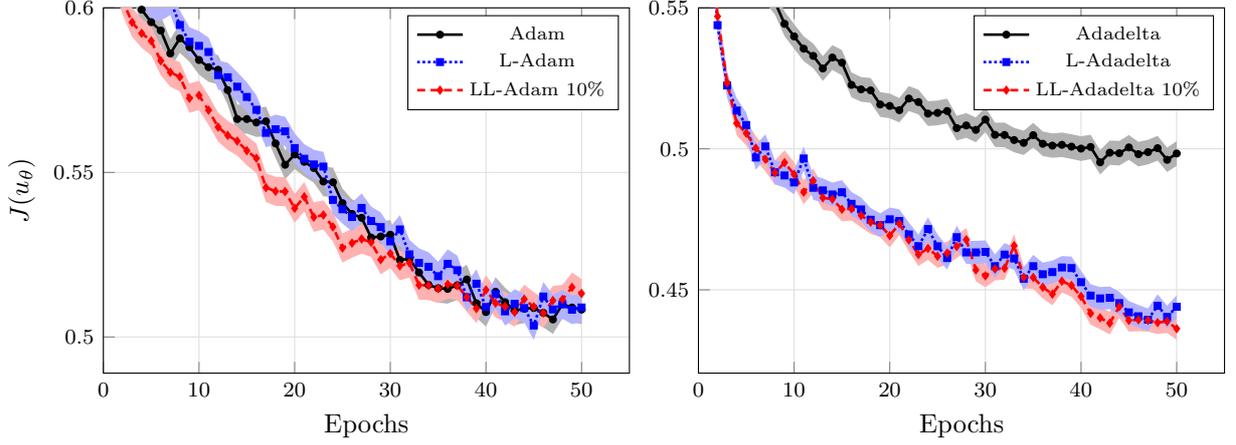
\begin{figure}
\centering

\begin{tikzpicture}
	\begin{groupplot}[group style={
					group size=2 by 1,
					horizontal sep=6ex,
					x descriptions at=edge bottom,
					group name=plots,
					},
			  xlabel=Epochs,
			  ytick pos=left,
			  height=0.25\textheight,
			  width=0.5\textwidth,
			  every axis plot/.append style={line width=1pt,mark size=1pt,mark options={solid}},
			  xmin=0,
			  grid=both,
			  minor grid style={gray!25},
			  major grid style={gray!25},
			  legend style={font=\scriptsize},
			  ]

		\nextgroupplot[ylabel=$J(u_\theta)$, ymax=0.6]
\addplot[color=black, mark=*]
	table[x=time,y=f,col sep=comma]{\base/fishing/layers_adam_N10/0_v_loss.csv};
\addlegendentry{Adam};
	
\addplot[color=blue, mark=square*, style=densely dotted]
	table[x=time,y=f,col sep=comma]{\base/fishing/layers_adam_N10/4_v_loss.csv};
\addlegendentry{L-Adam};

\addplot[color=red, mark=diamond*, style=densely dashed]
	table[x=time,y=f,col sep=comma]{\base/fishing/layers_adam_N10/1_v_loss.csv};
\addlegendentry{LL-Adam 10\%};
	
	\addplot[draw=none, name path=0_plus] table[x=time,y=f_plus,col sep=comma]{\base/fishing/layers_adam_N10/0_v_loss.csv};
	\addplot[draw=none, name path=0_minus] table[x=time,y=f_minus,col sep=comma]{\base/fishing/layers_adam_N10/0_v_loss.csv};
	\addplot[black!30] fill between[of=0_minus and 0_plus];

	\addplot[draw=none, name path=1_plus] table[x=time,y=f_plus,col sep=comma]{\base/fishing/layers_adam_N10/1_v_loss.csv};
	\addplot[draw=none, name path=1_minus] table[x=time,y=f_minus,col sep=comma]{\base/fishing/layers_adam_N10/1_v_loss.csv};
	\addplot[red!30] fill between[of=1_minus and 1_plus];
	
	\addplot[draw=none, name path=4_plus] table[x=time,y=f_plus,col sep=comma]{\base/fishing/layers_adam_N10/4_v_loss.csv};
	\addplot[draw=none, name path=4_minus] table[x=time,y=f_minus,col sep=comma]{\base/fishing/layers_adam_N10/4_v_loss.csv};
	\addplot[blue!30] fill between[of=4_minus and 4_plus];

		\nextgroupplot[ymax=0.55]
\addplot[color=black, mark=*]
	table[x=time,y=f,col sep=comma]{\base/fishing/layers_adadelta_N10/0_v_loss.csv};
\addlegendentry{Adadelta};
	
\addplot[color=blue, mark=square*, style=densely dotted]
	table[x=time,y=f,col sep=comma]{\base/fishing/layers_adadelta_N10/2_v_loss.csv};
\addlegendentry{L-Adadelta};

\addplot[color=red, mark=diamond*, style=densely dashed]
	table[x=time,y=f,col sep=comma]{\base/fishing/layers_adadelta_N10/1_v_loss.csv};
\addlegendentry{LL-Adadelta 10\%};
	
	\addplot[draw=none, name path=0_plus] table[x=time,y=f_plus,col sep=comma]{\base/fishing/layers_adadelta_N10/0_v_loss.csv};
	\addplot[draw=none, name path=0_minus] table[x=time,y=f_minus,col sep=comma]{\base/fishing/layers_adadelta_N10/0_v_loss.csv};
	\addplot[black!30] fill between[of=0_minus and 0_plus];

	\addplot[draw=none, name path=1_plus] table[x=time,y=f_plus,col sep=comma]{\base/fishing/layers_adadelta_N10/1_v_loss.csv};
	\addplot[draw=none, name path=1_minus] table[x=time,y=f_minus,col sep=comma]{\base/fishing/layers_adadelta_N10/1_v_loss.csv};
	\addplot[red!30] fill between[of=1_minus and 1_plus];
	
	\addplot[draw=none, name path=4_plus] table[x=time,y=f_plus,col sep=comma]{\base/fishing/layers_adadelta_N10/2_v_loss.csv};
	\addplot[draw=none, name path=4_minus] table[x=time,y=f_minus,col sep=comma]{\base/fishing/layers_adadelta_N10/2_v_loss.csv};
	\addplot[blue!30] fill between[of=4_minus and 4_plus];
	
	\end{groupplot}
\end{tikzpicture}

\caption{Training of the fishing problem with multiple controls with $N=10$. The schedules are $\gamma_n=2\e-3$ and $\sigma_n=2\e-3$ for Adam and $\gamma_n=5\e-1$ and $\sigma_n=5\e-3$ for Adadelta, for epochs 0 to 40 $\gamma_n$ is divided by 10 and $\sigma_n$ is set to 0 beyond.}
\label{fig:fishing:layers}
\end{figure}

The results are given in Figure \ref{fig:fishing:1} for the Adam optimizer with an increasing number of Euler-Maruyama steps and with one single control, in Figure \ref{fig:fishing:2} for the RMSprop and L-Adadelta optimizers and with one single control and in Figure \ref{fig:fishing:layers} for the training with multiple neural networks.

\begin{figure}
\centering

\begin{tikzpicture}
	\begin{groupplot}[group style={
					group size=2 by 1,
					horizontal sep=3ex,
					x descriptions at=edge bottom,
					group name=plots,
					},
			  xlabel=Time$\times 100$,
			  ytick pos=left,
			  height=0.2\textheight,
			  width=0.5\textwidth,
			  every axis plot/.append style={line width=2pt,mark size=0.pt,mark options={solid}},
			  xmin=0,
			  grid=both,
			  minor grid style={gray!25},
			  major grid style={gray!25},
			  ]

		\nextgroupplot[ylabel=Fish biomass]
\addplot[color=black, mark=*]
	table[x=time,y=f0,col sep=comma]{\base/fishing/trajs/traj_example.csv};
\addplot[color=blue, mark=*, style=densely dotted]
	table[x=time,y=f1,col sep=comma]{\base/fishing/trajs/traj_example.csv};
\addplot[color=red, mark=diamond*, style=densely dashed]
	table[x=time,y=f2,col sep=comma]{\base/fishing/trajs/traj_example.csv};
\addplot[color=olive, mark=square*, style=densely dashdotted]
	table[x=time,y=f3,col sep=comma]{\base/fishing/trajs/traj_example.csv};
\addplot[color=Plum, mark=pentagon*, style=densely dashdotted]
	table[x=time,y=f4,col sep=comma]{\base/fishing/trajs/traj_example.csv};

		\nextgroupplot[ylabel=Controlled Fishing, yticklabel pos=right, ylabel near ticks]
\addplot[color=black, mark=*]
	table[x=time,y=f5,col sep=comma]{\base/fishing/trajs/traj_example.csv};
\addplot[color=blue, mark=*, style=densely dotted]
	table[x=time,y=f6,col sep=comma]{\base/fishing/trajs/traj_example.csv};
\addplot[color=red, mark=diamond*, style=densely dashed]
	table[x=time,y=f7,col sep=comma]{\base/fishing/trajs/traj_example.csv};
\addplot[color=olive, mark=square*, style=densely dashdotted]
	table[x=time,y=f8,col sep=comma]{\base/fishing/trajs/traj_example.csv};
\addplot[color=Plum, mark=pentagon*, style=densely dashdotted]
	table[x=time,y=f9,col sep=comma]{\base/fishing/trajs/traj_example.csv};
	
	\end{groupplot}
\end{tikzpicture}

\caption{Example of a trajectory of $X_t \in \mathbb{R}^5$ along with the controlled fishing $u_t$ with $N=100$. We recall that the objective biomass is $\mathcal{X}_t \equiv \mathbf{1}$.}
\label{fig:fishing:traj}
\end{figure}
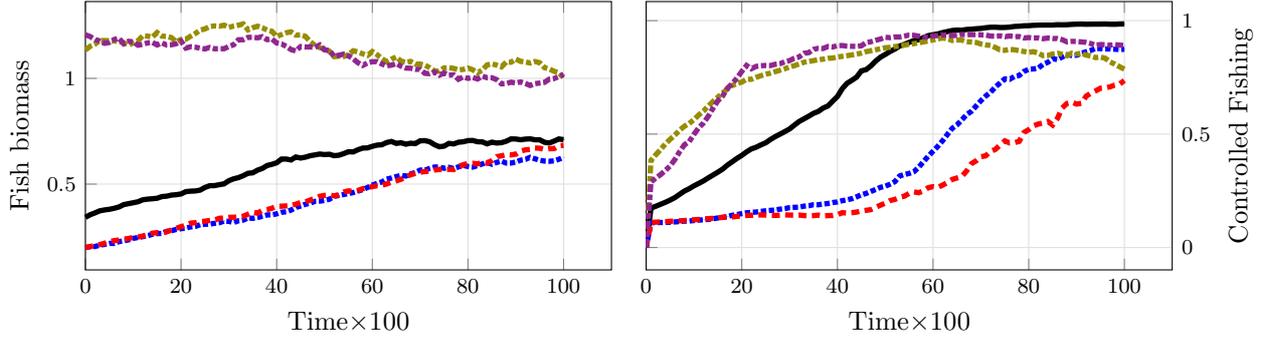

\begin{figure}
\centering

\begin{tikzpicture}
	\begin{groupplot}[group style={
					group size=3 by 1,
					horizontal sep=5ex,
					x descriptions at=edge bottom,
					group name=plots,
					},
			  ytick pos=left,
			  height=0.25\textheight,
			  width=0.35\textwidth,
			  every axis plot/.append style={line width=1pt,mark size=1.pt,mark options={solid}},
			  xmin=5,
			  xmax=50,
			  ymax=0.6,
			  grid=both,
			  minor grid style={gray!25},
			  major grid style={gray!25},
			  ]

  		\nextgroupplot[ylabel=$J(u_\theta)$,ymax=0.5,xmin=10,]
\addplot[color=black, mark=*]
	table[x=time,y=f,col sep=comma]{\base/fishing/adam_N20/0_v_loss.csv};
	\addplot[draw=none, name path=0_plus] table[x=time,y=f_plus,col sep=comma]{\base/fishing/adam_N20/0_v_loss.csv};
	\addplot[draw=none, name path=0_minus] table[x=time,y=f_minus,col sep=comma]{\base/fishing/adam_N20/0_v_loss.csv};
	\addplot[black!30] fill between[of=0_minus and 0_plus];
	
\addplot[color=blue, mark=square*, style=densely dotted]
	table[x=time,y=f,col sep=comma]{\base/fishing/adam_N20/1_v_loss.csv};
	\addplot[draw=none, name path=1_plus] table[x=time,y=f_plus,col sep=comma]{\base/fishing/adam_N20/1_v_loss.csv};
	\addplot[draw=none, name path=1_minus] table[x=time,y=f_minus,col sep=comma]{\base/fishing/adam_N20/1_v_loss.csv};
	\addplot[blue!30] fill between[of=1_minus and 1_plus];

		\nextgroupplot[xlabel=Epochs]
\addplot[color=black, mark=*] %
	table[x=time,y=f,col sep=comma]{\base/fishing/adam_N50/0_v_loss.csv};
	\addplot[draw=none, name path=0_plus] table[x=time,y=f_plus,col sep=comma]{\base/fishing/adam_N50/0_v_loss.csv};
	\addplot[draw=none, name path=0_minus] table[x=time,y=f_minus,col sep=comma]{\base/fishing/adam_N50/0_v_loss.csv};
	\addplot[black!30] fill between[of=0_minus and 0_plus];

\addplot[color=blue, mark=square*, style=densely dotted] %
	table[x=time,y=f,col sep=comma]{\base/fishing/adam_N50/1_v_loss.csv};
	\addplot[draw=none, name path=1_plus] table[x=time,y=f_plus,col sep=comma]{\base/fishing/adam_N50/1_v_loss.csv};
	\addplot[draw=none, name path=1_minus] table[x=time,y=f_minus,col sep=comma]{\base/fishing/adam_N50/1_v_loss.csv};
	\addplot[blue!30] fill between[of=1_minus and 1_plus];

		\nextgroupplot[]
\addplot[color=black, mark=*]
	table[x=time,y=f,col sep=comma]{\base/fishing/adam_N100/0_v_loss.csv};
\addlegendentry{Adam};

\addplot[color=blue, mark=square*, style=densely dotted] %
	table[x=time,y=f,col sep=comma]{\base/fishing/adam_N100/1_v_loss.csv};
\addlegendentry{L-Adam};

	\addplot[draw=none, name path=0_plus] table[x=time,y=f_plus,col sep=comma]{\base/fishing/adam_N100/0_v_loss.csv};
	\addplot[draw=none, name path=0_minus] table[x=time,y=f_minus,col sep=comma]{\base/fishing/adam_N100/0_v_loss.csv};
	\addplot[black!30] fill between[of=0_minus and 0_plus];

	\addplot[draw=none, name path=1_plus] table[x=time,y=f_plus,col sep=comma]{\base/fishing/adam_N100/1_v_loss.csv};
	\addplot[draw=none, name path=1_minus] table[x=time,y=f_minus,col sep=comma]{\base/fishing/adam_N100/1_v_loss.csv};
	\addplot[blue!30] fill between[of=1_minus and 1_plus];

\end{groupplot}
\end{tikzpicture}

\begin{tikzpicture}
\begin{groupplot}[group style={
					group size=3 by 1,
					horizontal sep=5ex,
					x descriptions at=edge bottom,
					group name=plots,
					},
			  ytick pos=left,
			  height=0.15\textheight,
			  width=0.35\textwidth,
			  every axis plot/.append style={line width=1pt,mark size=1.pt,mark options={solid}},
			  xmin=30,
			  xmax=50,
			  ymax=0.41,
			  grid=both,
			  minor grid style={gray!25},
			  major grid style={gray!25},
			  ]

  		\nextgroupplot[ylabel=Zoom,]
\addplot[color=black, mark=*]
	table[x=time,y=f,col sep=comma]{\base/fishing/adam_N20/0_v_loss.csv};
	\addplot[draw=none, name path=0_plus] table[x=time,y=f_plus,col sep=comma]{\base/fishing/adam_N20/0_v_loss.csv};
	\addplot[draw=none, name path=0_minus] table[x=time,y=f_minus,col sep=comma]{\base/fishing/adam_N20/0_v_loss.csv};
	\addplot[black!30] fill between[of=0_minus and 0_plus];
	
\addplot[color=blue, mark=square*, style=densely dotted]
	table[x=time,y=f,col sep=comma]{\base/fishing/adam_N20/1_v_loss.csv};
	\addplot[draw=none, name path=1_plus] table[x=time,y=f_plus,col sep=comma]{\base/fishing/adam_N20/1_v_loss.csv};
	\addplot[draw=none, name path=1_minus] table[x=time,y=f_minus,col sep=comma]{\base/fishing/adam_N20/1_v_loss.csv};
	\addplot[blue!50,opacity=0.5,] fill between[of=1_minus and 1_plus];

		\nextgroupplot[]
\addplot[color=black, mark=*] %
	table[x=time,y=f,col sep=comma]{\base/fishing/adam_N50/0_v_loss.csv};
	\addplot[draw=none, name path=0_plus] table[x=time,y=f_plus,col sep=comma]{\base/fishing/adam_N50/0_v_loss.csv};
	\addplot[draw=none, name path=0_minus] table[x=time,y=f_minus,col sep=comma]{\base/fishing/adam_N50/0_v_loss.csv};
	\addplot[black!30] fill between[of=0_minus and 0_plus];

\addplot[color=blue, mark=square*, style=densely dotted] %
	table[x=time,y=f,col sep=comma]{\base/fishing/adam_N50/1_v_loss.csv};
	\addplot[draw=none, name path=1_plus] table[x=time,y=f_plus,col sep=comma]{\base/fishing/adam_N50/1_v_loss.csv};
	\addplot[draw=none, name path=1_minus] table[x=time,y=f_minus,col sep=comma]{\base/fishing/adam_N50/1_v_loss.csv};
	\addplot[blue!50,opacity=0.5,] fill between[of=1_minus and 1_plus,];

		\nextgroupplot[ymax=0.42,]
\addplot[color=black, mark=*]
	table[x=time,y=f,col sep=comma]{\base/fishing/adam_N100/0_v_loss.csv};

\addplot[color=blue, mark=square*, style=densely dotted] %
	table[x=time,y=f,col sep=comma]{\base/fishing/adam_N100/1_v_loss.csv};

	\addplot[draw=none, name path=0_plus] table[x=time,y=f_plus,col sep=comma]{\base/fishing/adam_N100/0_v_loss.csv};
	\addplot[draw=none, name path=0_minus] table[x=time,y=f_minus,col sep=comma]{\base/fishing/adam_N100/0_v_loss.csv};
	\addplot[black!30] fill between[of=0_minus and 0_plus];

	\addplot[draw=none, name path=1_plus] table[x=time,y=f_plus,col sep=comma]{\base/fishing/adam_N100/1_v_loss.csv};
	\addplot[draw=none, name path=1_minus] table[x=time,y=f_minus,col sep=comma]{\base/fishing/adam_N100/1_v_loss.csv};
	\addplot[blue!50,opacity=0.5,] fill between[of=1_minus and 1_plus];

\end{groupplot}

\end{tikzpicture}

\caption{Comparison of Adam et L-Adam algorithms during the training for the fishing control problem with $N=20,50,100$ respectively. The schedules are $\gamma_n=2\e-3$ and $\sigma_n=1\e-3$ ($5\e-3$ for $N=100$) for epochs 0 to 40 and $\gamma_n=2\e-4$ and $\sigma_n=0$ beyond. At the end of each epoch, $J$ is estimated over $50 \times 512$ trajectories. A zoom on the last epochs is given.}
\label{fig:fishing:1}
\end{figure}
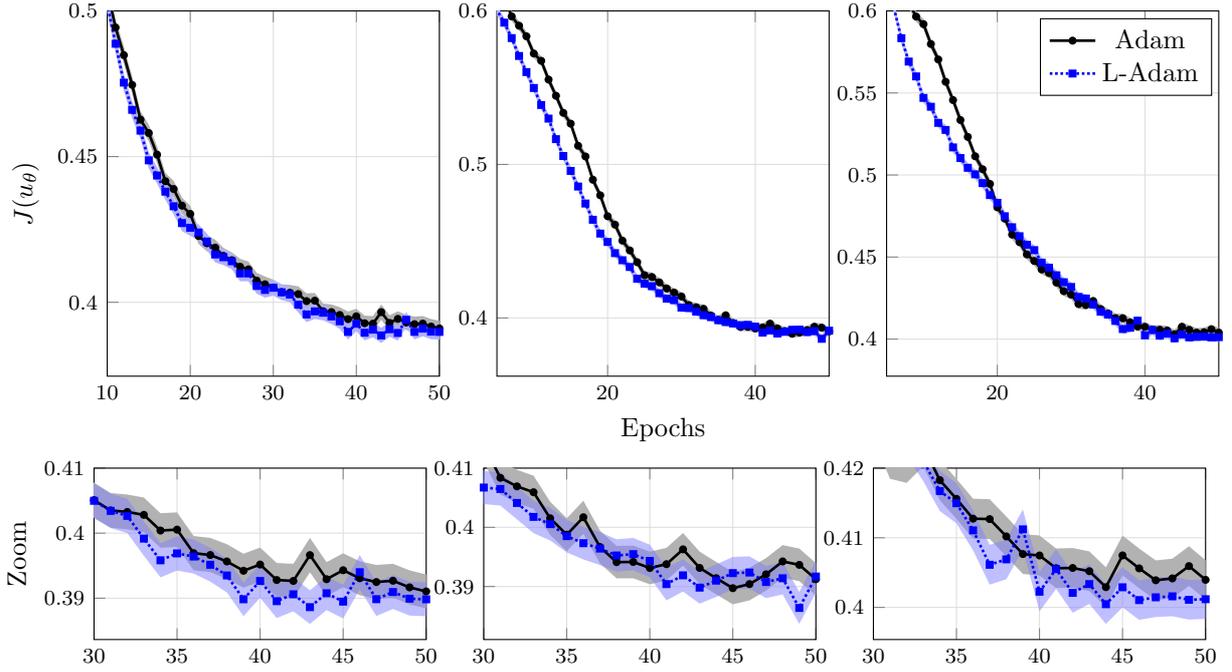

\section{Deep hedging}

We consider the problem of hedging portfolio of derivatives as a SOC problem as in \cite{buehler2019}. We aim to replicate a $\mathcal{F}_T$-measurable payoff $Z$ defined on some portfolio $S_t \in \mathbb{R}^{d_1}$ by trading (at least some of) the assets contained in $S_t$ at times $(t_k)$. The control is given by $u_t \in \mathbb{R}^{d_1}$ representing the amount held for each asset. The objective is
\begin{equation}
\label{eq:hedging:1}
J(u) = \nu \left( -Z + \sum_{k=0}^{N-1} \langle u_{t_k}, S_{t_{k+1}}-S_{t_k} \rangle - \sum_{k=0}^N \langle c_{tr} , S_{t_k} \ast |u_{t_k} - u_{t_{k-1}}| \rangle \right)
\end{equation}
where $\nu : L^1(\Omega) \to \mathbb{R}$ is a convex risk measure (see \cite[Definition 3.1]{buehler2019}), $c_{tr} \in \mathbb{R}^{d_1}$ represents proportional transaction costs and we fix $u_{t_{-1}} = u_{t_{N}} = 0$, implying full liquidation in $T$. We furthermore assume that $\nu$ can be written as
\begin{equation}
\nu(X) = \inf_{w \in \mathbb{R}} \left( w + \mathbb{E}[ \ell(-X -w)] \right)
\end{equation}
where the loss function $\ell : \mathbb{R} \to \mathbb{R}$ is continuous, non-decreasing and convex. This is the case in particular for the entropic risk measure where $\ell(x)=-\exp(-\lambda x)$ and the conditional value at risk measure where $\ell(x)=(1-\alpha)^{-1} \max(x,0)$. Then  \eqref{eq:hedging:1} can be rewritten as
\begin{equation}
\inf_{u, w} J(u,w) := \mathbb{E} \left[ w + \ell \left(Z - \sum_{k=0}^{N-1} \langle u_{t_k}, S_{t_{k+1}}-S_{t_k} \rangle + \sum_{k=0}^N \langle c_{tr} , S_{t_k} \ast |u_{t_k} - u_{t_{k-1}}| \rangle -w \right) \right].
\end{equation}

In the numerical experiments, we analyse the problem of hedging in a Heston model as described in \cite[Section 5]{buehler2019}.
For even $d_1$, we consider $d_1' := d_1/2$ independent Heston models where the price and volatility processes are described by the following SDEs for $1 \le i \le d_1'$:
\begin{align}
& dS^{1,i}_t = \sqrt{V_t^i} S^{1,i}_t dB^i_t, \quad S_0^{1,i} = s_0^i, \\
& dV_t^i = a^i(b^i - V_t^i) dt + \eta^i \sqrt{V_t^i} dW^i_t, \quad V_0^i = v_0^i,
\end{align}
where $a, b, \eta, s_0, v_i \in (\mathbb{R}^+)^{d_1'}$ and for each $1 \le i \le d_1'$, $B^i$ and $W^i$ are standard Brownian motions with correlation $\rho^i \in [-1,1]$.
The volatility $V$ itself is not tradable directly but only through options on variance modelled by the following variance swap:
\begin{align}
& S^{2,i}_t := \mathbb{E} \left[ \int_0^T V_s^i ds \Big| \ \mathcal{F}_t \right] = \int_0^t V_s^i ds + L^i(t,V_t^i), \\
& L^i(t,v) := \frac{v-b^i}{a^i} \left( 1 - e^{a^i(T-t)} \right) + b^i(T-t).
\end{align}
The payoff is given by
$$ Z = \sum_{i=1}^{d_1'} \big( S^{1,i}_T - K^i \big)_+ $$
where $K \in (\mathbb{R}^+)^{d_1'}$.
We consider the convex risk measure associated to the value-at-risk i.e. associated to the loss function
$$ \ell(x) = \frac{1}{1-\alpha} \max(x,0) .$$

In the experiments we choose
\begin{align}
& d_1'=5, \ T=1, \ a = \mathbf{1}, \ b = 0.04 \ast \mathbf{1}, \ \eta = 2 \ast \mathbf{1}, \ \rho=-0.7*\mathbf{1}, \ \alpha=0.9,\\
& s_0=K=\mathbf{1}, \ v_0 = 0.1 \ast \mathbf{1}, \ c_{\text{tr}} = 5\e-4 \ast \mathbf{1}.
\end{align}

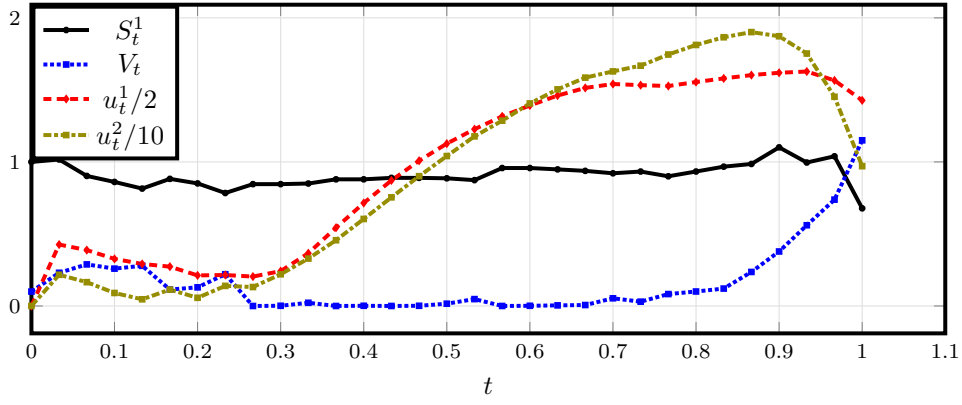
\begin{figure}
\centering

\begin{tikzpicture}
\begin{axis}[
	xlabel=$t$,
	xmin=0,
	grid=both,
	minor grid style={gray!25},
	major grid style={gray!25},
	legend style={at={(0,1)},anchor=north west},
	width=0.8\linewidth,
	height=0.2\paperheight,
	line width=1.5pt,
	mark size=0.5pt,
	mark options={solid},
	]
\addplot[color=black, mark=*]
	table[x expr=\thisrow{time}/30,y=f0,col sep=comma]{\base/deep_hedging/trajs/traj_example.csv};
\addlegendentry{$S^1_t$};

\addplot[color=blue, mark=square*, style=densely dotted]
	table[x expr=\thisrow{time}/30,y=f5,col sep=comma]{\base/deep_hedging/trajs/traj_example.csv};
\addlegendentry{$V_t$};

\addplot[color=red, mark=diamond*, style=densely dashed]
	table[x expr=\thisrow{time}/30,y expr=\thisrow{f10}/2,col sep=comma]{\base/deep_hedging/trajs/traj_example.csv};
\addlegendentry{$u^1_t/2$};
	
\addplot[color=olive, mark=square*, style=densely dashdotted]
	table[x expr=\thisrow{time}/30,y expr=\thisrow{f15}/10,col sep=comma]{\base/deep_hedging/trajs/traj_example.csv};
\addlegendentry{$u^2_t/10$};

\end{axis}
\end{tikzpicture}

\caption{Example of trajectory for the deep hedging problem with $N=30$.}
\label{fig:deep_hedging:traj}
\end{figure}

Each control $u_\theta$ is given by a feedforward neural network with two hidden layers with 32 units each and with ReLU activation while the output layer has ReLU activation too in order to forbid short-selling. As recommended in \cite{buehler2019}, since transaction costs are involved the control $u_\theta$ at time $t_{k}$ is a function of $\log(S^1_{t_k})$, $V_{t_k}$ and $u_{t_{k-1}}$.
An example of controlled trajectory showing only one of the five Heston models is plotted in Figure \ref{fig:deep_hedging:traj}.

\begin{figure}
\centering

\begin{tikzpicture}
	\begin{groupplot}[group style={
					group size=3 by 1,
					horizontal sep=4ex,
					x descriptions at=edge bottom,
					group name=plots,
					},
			  ytick pos=left,
			  height=0.25\textheight,
			  width=0.35\textwidth,
			  every axis plot/.append style={line width=1pt,mark size=0.2pt,mark options={solid}},
			  xmin=0,xmax=50,
			  grid=both,
			  minor grid style={gray!25},
			  major grid style={gray!25},
			  ]

		\nextgroupplot[ylabel=$J(u_\theta)$, ymax=0.8]
\addplot[color=black, mark=*]
	table[x=time,y=f,col sep=comma]{\base/deep_hedging/adam_N30/0_v_loss.csv};
\addlegendentry{Adam};

\addplot[color=blue, mark=square*, style=densely dotted]
	table[x=time,y=f,col sep=comma]{\base/deep_hedging/adam_N30/1_v_loss.csv};
\addlegendentry{L-Adam};

	\addplot[draw=none, name path=0_plus] table[x=time,y=f_plus,col sep=comma]{\base/deep_hedging/adam_N30/0_v_loss.csv};
	\addplot[draw=none, name path=0_minus] table[x=time,y=f_minus,col sep=comma]{\base/deep_hedging/adam_N30/0_v_loss.csv};
	\addplot[black!30] fill between[of=0_minus and 0_plus];

	\addplot[draw=none, name path=1_plus] table[x=time,y=f_plus,col sep=comma]{\base/deep_hedging/adam_N30/1_v_loss.csv};
	\addplot[draw=none, name path=1_minus] table[x=time,y=f_minus,col sep=comma]{\base/deep_hedging/adam_N30/1_v_loss.csv};
	\addplot[blue!30] fill between[of=1_minus and 1_plus];

		\nextgroupplot[xmin=10,xmax=100,ymax=4,xlabel=Epochs]
\addplot[color=black, mark=*]
	table[x=time,y=f,col sep=comma]{\base/deep_hedging/adam_N50/0_v_loss.csv};
\addlegendentry{Adam};

\addplot[color=blue, mark=square*, style=densely dotted]
	table[x=time,y=f,col sep=comma]{\base/deep_hedging/adam_N50/1_v_loss.csv};
\addlegendentry{L-Adam};

	\addplot[draw=none, name path=0_plus] table[x=time,y=f_plus,col sep=comma]{\base/deep_hedging/adam_N50/0_v_loss.csv};
	\addplot[draw=none, name path=0_minus] table[x=time,y=f_minus,col sep=comma]{\base/deep_hedging/adam_N50/0_v_loss.csv};
	\addplot[black!30] fill between[of=0_minus and 0_plus];

	\addplot[draw=none, name path=1_plus] table[x=time,y=f_plus,col sep=comma]{\base/deep_hedging/adam_N50/1_v_loss.csv};
	\addplot[draw=none, name path=1_minus] table[x=time,y=f_minus,col sep=comma]{\base/deep_hedging/adam_N50/1_v_loss.csv};
	\addplot[blue!30] fill between[of=1_minus and 1_plus];

		\nextgroupplot[ymax=1]
\addplot[color=black, mark=*]
	table[x=time,y=f,col sep=comma]{\base/deep_hedging/adadelta_N50/0_v_loss.csv};
\addlegendentry{Adadelta};

\addplot[color=blue, mark=square*, style=densely dotted]
	table[x=time,y=f,col sep=comma]{\base/deep_hedging/adadelta_N50/1_v_loss.csv};
\addlegendentry{L-Adadelta};

	\addplot[draw=none, name path=0_plus] table[x=time,y=f_plus,col sep=comma]{\base/deep_hedging/adadelta_N50/0_v_loss.csv};
	\addplot[draw=none, name path=0_minus] table[x=time,y=f_minus,col sep=comma]{\base/deep_hedging/adadelta_N50/0_v_loss.csv};
	\addplot[black!30] fill between[of=0_minus and 0_plus];

	\addplot[draw=none, name path=1_plus] table[x=time,y=f_plus,col sep=comma]{\base/deep_hedging/adadelta_N50/1_v_loss.csv};
	\addplot[draw=none, name path=1_minus] table[x=time,y=f_minus,col sep=comma]{\base/deep_hedging/adadelta_N50/1_v_loss.csv};
	\addplot[blue!30] fill between[of=1_minus and 1_plus];
	
	\end{groupplot}
\end{tikzpicture}

\caption{Comparison of algorithms during the training for the deep hedging control problem with $N=30,50,50$ respectively. The schedules are $\gamma_n=2\e-3$ ($5\e-1$) and $\sigma_n=2\e-3$ ($5\e-3$) for Adam (resp. Adadelta) for epochs 0 to 80 and $\gamma_n$ is divided by 10 and $\sigma_n$ is set to 0 beyond.}
\label{fig:deep_hedging:1}
\end{figure}
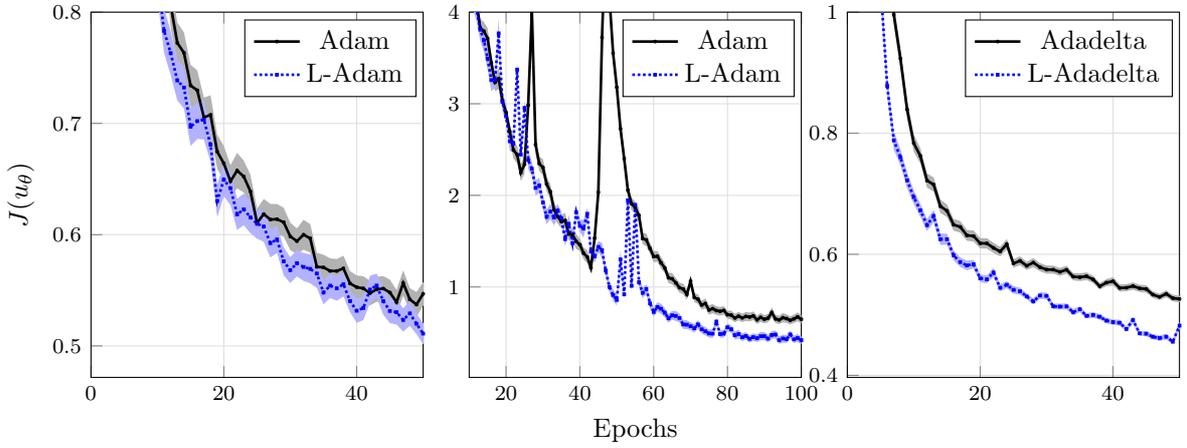

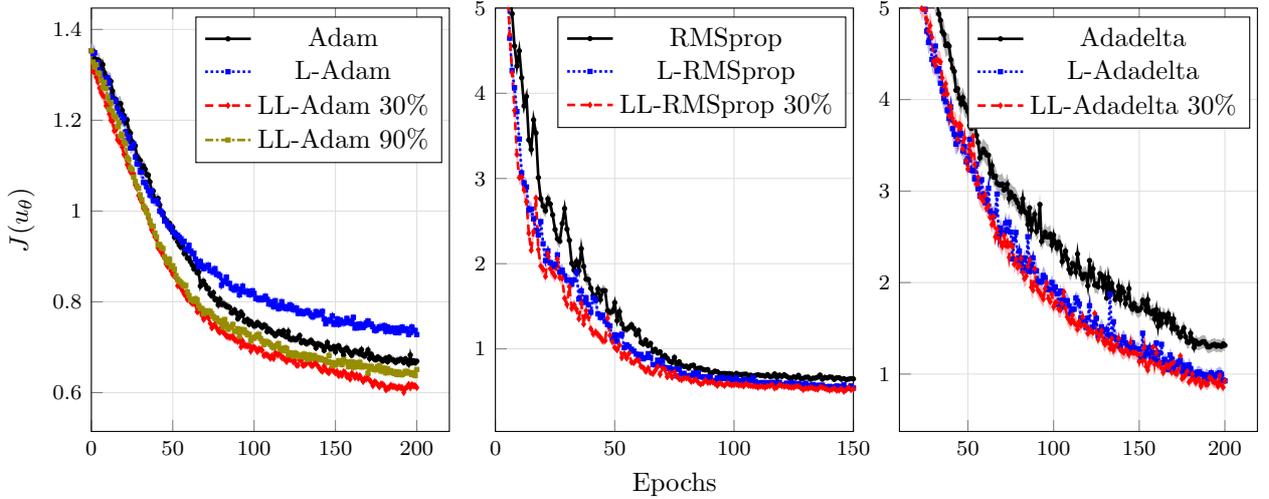
\begin{figure}
\centering

\begin{tikzpicture}
	\begin{groupplot}[group style={
					group size=3 by 1,
					horizontal sep=4ex,
					x descriptions at=edge bottom,
					group name=plots,
					},
			  ytick pos=left,
			  height=0.28\textheight,
			  width=0.37\textwidth,
			  every axis plot/.append style={line width=1pt,mark size=0.5pt,mark options={solid}},
			  xmin=0,
			  grid=both,
			  minor grid style={gray!25},
			  major grid style={gray!25},
			  ]

		\nextgroupplot[ylabel=$J(u_\theta)$,]
\addplot[color=black, mark=*]
	table[x=time,y=f,col sep=comma]{\base/deep_hedging/layers_adam_N10_7/0_v_loss.csv};
\addlegendentry{Adam};

\addplot[color=blue, mark=square*, style=densely dotted]
	table[x=time,y=f,col sep=comma]{\base/deep_hedging/layers_adam_N10_7/1_v_loss.csv};
\addlegendentry{L-Adam};

\addplot[color=red, mark=diamond*, style=densely dashed] %
	table[x=time,y=f,col sep=comma]{\base/deep_hedging/layers_adam_N10_7/3_v_loss.csv};
\addlegendentry{LL-Adam 30\%};

\addplot[color=olive, mark=square*, style=densely dashdotted] %
	table[x=time,y=f,col sep=comma]{\base/deep_hedging/layers_adam_N10_7/4_v_loss.csv};
\addlegendentry{LL-Adam 90\%};

	\addplot[draw=none, name path=0_plus] table[x=time,y=f_plus,col sep=comma]{\base/deep_hedging/layers_adam_N10_7/0_v_loss.csv};
	\addplot[draw=none, name path=0_minus] table[x=time,y=f_minus,col sep=comma]{\base/deep_hedging/layers_adam_N10_7/0_v_loss.csv};
	\addplot[black!30] fill between[of=0_minus and 0_plus];

	\addplot[draw=none, name path=1_plus] table[x=time,y=f_plus,col sep=comma]{\base/deep_hedging/layers_adam_N10_7/1_v_loss.csv};
	\addplot[draw=none, name path=1_minus] table[x=time,y=f_minus,col sep=comma]{\base/deep_hedging/layers_adam_N10_7/1_v_loss.csv};
	\addplot[blue!30] fill between[of=1_minus and 1_plus];
	
	\addplot[draw=none, name path=3_plus] table[x=time,y=f_plus,col sep=comma]{\base/deep_hedging/layers_adam_N10_7/3_v_loss.csv};
	\addplot[draw=none, name path=3_minus] table[x=time,y=f_minus,col sep=comma]{\base/deep_hedging/layers_adam_N10_7/3_v_loss.csv};
	\addplot[red!30] fill between[of=3_minus and 3_plus];
	
	\addplot[draw=none, name path=4_plus] table[x=time,y=f_plus,col sep=comma]{\base/deep_hedging/layers_adam_N10_7/4_v_loss.csv};
	\addplot[draw=none, name path=4_minus] table[x=time,y=f_minus,col sep=comma]{\base/deep_hedging/layers_adam_N10_7/4_v_loss.csv};
	\addplot[olive!30] fill between[of=4_minus and 4_plus];

	\nextgroupplot[xlabel=Epochs,ymax=5,xmax=150,]
\addplot[color=black, mark=*]
	table[x=time,y=f,col sep=comma]{\base/deep_hedging/layers_rmsprop_N10_3/0_v_loss.csv};
\addlegendentry{RMSprop};

\addplot[color=blue, mark=square*, style=densely dotted]
	table[x=time,y=f,col sep=comma]{\base/deep_hedging/layers_rmsprop_N10_3/1_v_loss.csv};
\addlegendentry{L-RMSprop};

\addplot[color=red, mark=diamond*, style=densely dashed] %
	table[x=time,y=f,col sep=comma]{\base/deep_hedging/layers_rmsprop_N10_3/2_v_loss.csv};
\addlegendentry{LL-RMSprop 30\%};

	\addplot[draw=none, name path=0_plus] table[x=time,y=f_plus,col sep=comma]{\base/deep_hedging/layers_rmsprop_N10_3/0_v_loss.csv};
	\addplot[draw=none, name path=0_minus] table[x=time,y=f_minus,col sep=comma]{\base/deep_hedging/layers_rmsprop_N10_3/0_v_loss.csv};
	\addplot[black!30] fill between[of=0_minus and 0_plus];

	\addplot[draw=none, name path=1_plus] table[x=time,y=f_plus,col sep=comma]{\base/deep_hedging/layers_rmsprop_N10_3/1_v_loss.csv};
	\addplot[draw=none, name path=1_minus] table[x=time,y=f_minus,col sep=comma]{\base/deep_hedging/layers_rmsprop_N10_3/1_v_loss.csv};
	\addplot[blue!30] fill between[of=1_minus and 1_plus];
	
	\addplot[draw=none, name path=2_plus] table[x=time,y=f_plus,col sep=comma]{\base/deep_hedging/layers_rmsprop_N10_3/2_v_loss.csv};
	\addplot[draw=none, name path=2_minus] table[x=time,y=f_minus,col sep=comma]{\base/deep_hedging/layers_rmsprop_N10_3/2_v_loss.csv};
	\addplot[red!30] fill between[of=2_minus and 2_plus];

		\nextgroupplot[xmin=10,ymax=5,]
\addplot[color=black, mark=*]
	table[x=time,y=f,col sep=comma]{\base/deep_hedging/layers_adadelta_N10_3/0_v_loss.csv};
\addlegendentry{Adadelta};

\addplot[color=blue, mark=square*, style=densely dotted]
	table[x=time,y=f,col sep=comma]{\base/deep_hedging/layers_adadelta_N10_3/1_v_loss.csv};
\addlegendentry{L-Adadelta};

\addplot[color=red, mark=diamond*, style=densely dashed]
	table[x=time,y=f,col sep=comma]{\base/deep_hedging/layers_adadelta_N10_3/2_v_loss.csv};
\addlegendentry{LL-Adadelta 30\%};

	\addplot[draw=none, name path=0_plus] table[x=time,y=f_plus,col sep=comma]{\base/deep_hedging/layers_adadelta_N10_3/0_v_loss.csv};
	\addplot[draw=none, name path=0_minus] table[x=time,y=f_minus,col sep=comma]{\base/deep_hedging/layers_adadelta_N10_3/0_v_loss.csv};
	\addplot[black!30] fill between[of=0_minus and 0_plus];

	\addplot[draw=none, name path=1_plus] table[x=time,y=f_plus,col sep=comma]{\base/deep_hedging/layers_adadelta_N10_3/1_v_loss.csv};
	\addplot[draw=none, name path=1_minus] table[x=time,y=f_minus,col sep=comma]{\base/deep_hedging/layers_adadelta_N10_3/1_v_loss.csv};
	\addplot[blue!30] fill between[of=1_minus and 1_plus];
	
	\addplot[draw=none, name path=2_plus] table[x=time,y=f_plus,col sep=comma]{\base/deep_hedging/layers_adadelta_N10_3/2_v_loss.csv};
	\addplot[draw=none, name path=2_minus] table[x=time,y=f_minus,col sep=comma]{\base/deep_hedging/layers_adadelta_N10_3/2_v_loss.csv};
	\addplot[red!30] fill between[of=2_minus and 2_plus];

	\end{groupplot}
\end{tikzpicture}

\caption{Training of the deep hedging problem with multiple controls with $N=10$. The schedules are $\gamma_n=2\e-3$ ($5\e-1$) and $\sigma_n=2\e-3$ ($5\e-3$) for Adam and RMSprop (resp. Adadelta) for epochs 0 to 180 and $\gamma_n$ is divided by 10 and $\sigma_n$ is set to 0 beyond.}
\label{fig:deep_hedging:2}
\end{figure}

The results are given in Figure \ref{fig:deep_hedging:1} for the comparison of Langevin and non-Langevin algorithms with a single control and in Figure \ref{fig:deep_hedging:2} for the training with multiple controls.

\section{Oil drilling}

We consider the control problem in the management of natural resources applied to oil drilling introduced in \cite{goutte2018} and extended in \cite{gaigi2021}. The objective is for an oil driller, to balance the costs of extraction, storage in a volatile energy market. The oil price $P_t \in \mathbb{R}$ is assumed to be a Black-Scholes process:
\begin{equation}
dP_t = \mu P_t dt + \eta P_t dW_t.
\end{equation}
The control is given by $q_t = (q^v_t, q^s_t, q^{v,s}_t) \in \mathbb{R}^3$ where $q^v_t$ is the quantity of extracted oil immediately sold on the market per time unit, $q^s_t$ is the quantity of extracted oil that is stored per time unit, $q^{v,s}_t$ is the quantity of stored oil that is sold per time unit. The cumulated quantities of extracted and stored oil at time $t$ are respectively given by
\begin{equation}
E_t = \int_0^t (q^v_r + q^s_r) dr, \quad S_t = \int_0^t (q^s_r - q^{v,s}_r) dr.
\end{equation}
The extraction and storage prices are respectively given by
\begin{equation}
c_e(E_t) = \exp\left(\xi_e E_t\right), \quad c_s(S_t) = \exp\left(\xi_s S_t\right) - 1.
\end{equation}
The constraints on the control are the following:
\begin{equation}
\label{eq:q_constraints}
q^v_t, \ q^s_t, \ q^{v,s}_t \ge 0, \quad q^{v,s}_t \le q^S, \quad q^v_t + q^s_t \le K_0, \quad 0 \le S_t \le Q^S,
\end{equation}
where $q^S$, $K_0$ and $Q^S$ are operational bounds. The objective is
\begin{equation}
J(q) = - \mathbb{E} \left[ \int_0^T e^{-\rho r} U\Big( q^v_r P_r + q^{v,s}_r (1-\varepsilon)P_r - (q^v_r + q^s_r) c_e(E_r) - c_s(S_r)\Big) dr \right],
\end{equation}
where $U : \mathbb{R} \to \mathbb{R}$ is the utility function.

In the experiments we take
\begin{align}
& T=1, \ \mu = 0.01, \ \eta = 0.2, \ \rho = 0.01, \ \varepsilon = 0, \ K_0 = 5, \nonumber \\
& \xi_e = 1\e-2, \ \xi_s = 5\e-3, \ q^S = 10, \ P_0 = 1, \ U(x) = x.
\end{align}

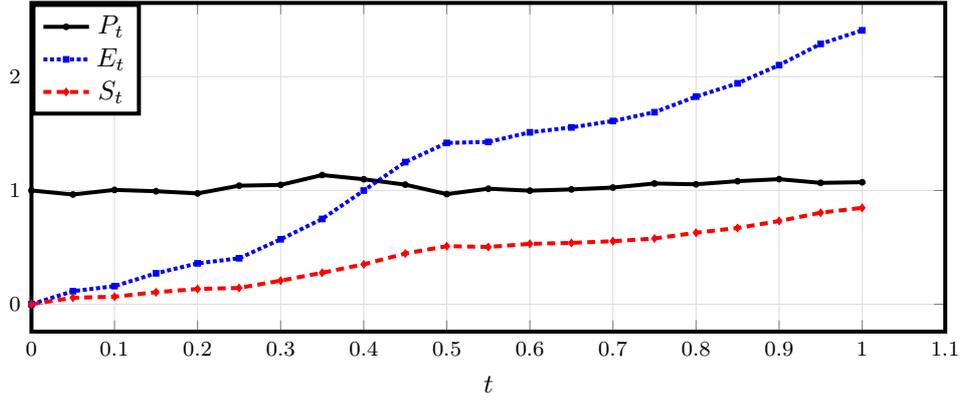
\begin{figure}
\centering

\begin{tikzpicture}
\begin{axis}[
	xlabel=$t$,
	xmin=0,
	grid=both,
	minor grid style={gray!25},
	major grid style={gray!25},
	legend style={at={(0,1)},anchor=north west},
	width=0.8\linewidth,
	height=0.2\paperheight,
	line width=1.5pt,
	mark size=0.5pt,
	mark options={solid},
	]
\addplot[color=black, mark=*]
	table[x expr=\thisrow{time}/20,y=f0,col sep=comma]{\base/oil_drilling/trajs/traj_example.csv};
\addlegendentry{$P_t$};

\addplot[color=blue, mark=square*, style=densely dotted]
	table[x expr=\thisrow{time}/20,y=f1,col sep=comma]{\base/oil_drilling/trajs/traj_example.csv};
\addlegendentry{$E_t$};

\addplot[color=red, mark=diamond*, style=densely dashed] %
	table[x expr=\thisrow{time}/20,y expr=\thisrow{f2}*1,col sep=comma]{\base/oil_drilling/trajs/traj_example.csv};
\addlegendentry{$S_t$};

\end{axis}
\end{tikzpicture}

\caption{Example of trajectory for the oil drilling problem with $N=20$.}
\label{fig:oil_drilling:traj}
\end{figure}

The control $q_t$ is given by a feedforward neural network with two hidden layers with 32 units and with $\relu$ activation while the output layer has several $\relu$ activations such that the constraints on $q$ \eqref{eq:q_constraints} are fulfilled\footnote{We remark that $\max(q,K) = K - \relu(-q+K)$.}. An example of controlled trajectory is given in Figure \ref{fig:oil_drilling:traj}.

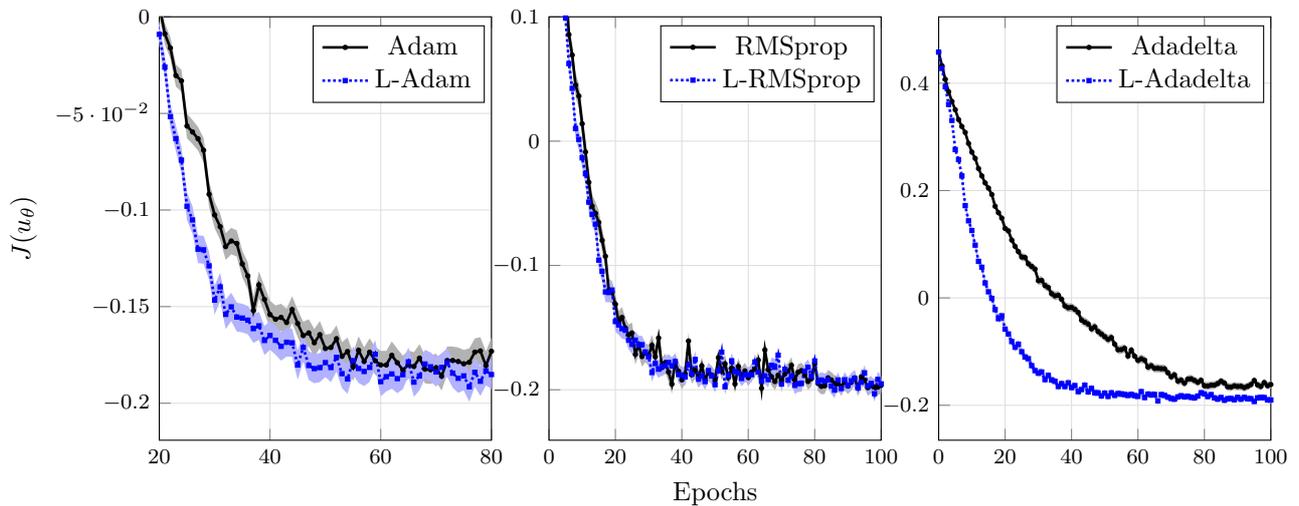
\begin{figure}
\centering

\begin{tikzpicture}
	\begin{groupplot}[group style={
					group size=3 by 1,
					horizontal sep=5ex,
					x descriptions at=edge bottom,
					group name=plots,
					},
			  ytick pos=left,
			  height=0.28\textheight,
			  width=0.35\textwidth,
			  every axis plot/.append style={line width=1pt,mark size=0.5pt,mark options={solid}},
			  xmin=0,xmax=100,
			  grid=both,
			  minor grid style={gray!25},
			  major grid style={gray!25},
			  ]

		\nextgroupplot[xmin=20,xmax=80,ymax=0,ylabel=$J(u_\theta)$]
\addplot[color=black, mark=*]
	table[x=time,y=f,col sep=comma]{\base/oil_drilling/adam_N50/0_v_loss.csv};
\addlegendentry{Adam};

\addplot[color=blue, mark=square*, style=densely dotted]
	table[x=time,y=f,col sep=comma]{\base/oil_drilling/adam_N50/1_v_loss.csv};
\addlegendentry{L-Adam};

	\addplot[draw=none, name path=0_plus] table[x=time,y=f_plus,col sep=comma]{\base/oil_drilling/adam_N50/0_v_loss.csv};
	\addplot[draw=none, name path=0_minus] table[x=time,y=f_minus,col sep=comma]{\base/oil_drilling/adam_N50/0_v_loss.csv};
	\addplot[black!30] fill between[of=0_minus and 0_plus];

	\addplot[draw=none, name path=1_plus] table[x=time,y=f_plus,col sep=comma]{\base/oil_drilling/adam_N50/1_v_loss.csv};
	\addplot[draw=none, name path=1_minus] table[x=time,y=f_minus,col sep=comma]{\base/oil_drilling/adam_N50/1_v_loss.csv};
	\addplot[blue!30] fill between[of=1_minus and 1_plus];

	\nextgroupplot[ymax=0.1,xlabel=Epochs,]
\addplot[color=black, mark=*]
	table[x=time,y=f,col sep=comma]{\base/oil_drilling/rmsprop_N50/0_v_loss.csv};
\addlegendentry{RMSprop};

\addplot[color=blue, mark=square*, style=densely dotted]
	table[x=time,y=f,col sep=comma]{\base/oil_drilling/rmsprop_N50/1_v_loss.csv};
\addlegendentry{L-RMSprop};

	\addplot[draw=none, name path=0_plus] table[x=time,y=f_plus,col sep=comma]{\base/oil_drilling/rmsprop_N50/0_v_loss.csv};
	\addplot[draw=none, name path=0_minus] table[x=time,y=f_minus,col sep=comma]{\base/oil_drilling/rmsprop_N50/0_v_loss.csv};
	\addplot[black!30] fill between[of=0_minus and 0_plus];

	\addplot[draw=none, name path=1_plus] table[x=time,y=f_plus,col sep=comma]{\base/oil_drilling/rmsprop_N50/1_v_loss.csv};
	\addplot[draw=none, name path=1_minus] table[x=time,y=f_minus,col sep=comma]{\base/oil_drilling/rmsprop_N50/1_v_loss.csv};
	\addplot[blue!30] fill between[of=1_minus and 1_plus];

		\nextgroupplot[]
\addplot[color=black, mark=*]
	table[x=time,y=f,col sep=comma]{\base/oil_drilling/adadelta_N50/0_v_loss.csv};
\addlegendentry{Adadelta};

\addplot[color=blue, mark=square*, style=densely dotted]
	table[x=time,y=f,col sep=comma]{\base/oil_drilling/adadelta_N50/1_v_loss.csv};
\addlegendentry{L-Adadelta};

	\addplot[draw=none, name path=0_plus] table[x=time,y=f_plus,col sep=comma]{\base/oil_drilling/adadelta_N50/0_v_loss.csv};
	\addplot[draw=none, name path=0_minus] table[x=time,y=f_minus,col sep=comma]{\base/oil_drilling/adadelta_N50/0_v_loss.csv};
	\addplot[black!30] fill between[of=0_minus and 0_plus];

	\addplot[draw=none, name path=1_plus] table[x=time,y=f_plus,col sep=comma]{\base/oil_drilling/adadelta_N50/1_v_loss.csv};
	\addplot[draw=none, name path=1_minus] table[x=time,y=f_minus,col sep=comma]{\base/oil_drilling/adadelta_N50/1_v_loss.csv};
	\addplot[blue!30] fill between[of=1_minus and 1_plus];
	
	\end{groupplot}
\end{tikzpicture}

\caption{Comparison of algorithms during the training for the deep hedging control problem with $N=50$. The schedules are $\gamma=2\e-3$ ($2\e-3$, $5\e-1$) and $\sigma=1\e-3$ ($2\e-3$, $5\e-3$) for Adam (resp. RMSprop, Adadelta) for epochs 0 to 60 (resp. 80, 80) and $\gamma$ is divided by 10 and $\sigma$ is set to 0 beyond.}
\label{fig:oil_drilling:1}
\end{figure}

The results are given in Figure \ref{fig:oil_drilling:1} for the comparison of Langevin and non-Langevin algorithms with a single control.
We do not display the results for the training with multiple controls however as we could not obtain satisfying results neither with Langevin nor-with non-Langevin methods.

\section{Comments on the numerical experiments}

We observe that in many cases and in various SOC problems, Langevin and Layer Langevin algorithms show improvement when compared with their respective non-Langevin counterparts, provided that $N$ is large enough, which is remarkable for randomized algorithms.
Langevin algorithms converge faster and/or toward a lower loss value.
This is particularly visible for the Adadelta method.
The gains are limited in some cases (see Figures \ref{fig:fishing:2} and \ref{fig:oil_drilling:1} for RMSprop) but still the optimization procedure is improved.

The gains for the L-RMSprop algorithm remain limited however. In particular, we did not observe any significant improvement for fishing SOC with multiple controls, for deep hedging SOC with a single control and for oil drilling SOC. We do not have explanation for this fact.

The gains brought by Langevin algorithm increase with the depth of the network as shown in Figures \ref{fig:fishing:1} and \ref{fig:deep_hedging:1}.
However, contrary to \cite{bras2022}, we did not observe overwhelming gains as $N$ becomes very large.
We believe that this is due (in part) to the particular structure of the deep SOC problem where the same control is repeated all along the trajectory.

As for SOC with multiple controls, we observe that Layer Langevin algorithms with a small number of Langevin layers (10\%-30\%) generally outperforms Vanilla Langevin methods while Vanilla Langevin may bring limited gains or be less efficient than the standard non-Langevin methods, see Figures \ref{fig:fishing:layers} and \ref{fig:deep_hedging:2} for Adam.

\section*{Acknowledgements}

The authors thank Idris Kharroubi for helpful discussions.


\newcommand{\etalchar}[1]{$^{#1}$}

\end{document}